\documentclass[twocolumn]{aastex62}

\usepackage{ulem}
\usepackage{amsmath}
\usepackage{color}
\begin{document}

\title{Three-Dimensional Simulation of Double-Detonations in the
  Double-Degenerate Model for Type Ia Supernovae and Interaction of
  Ejecta with a Surviving White Dwarf Companion}

\correspondingauthor{Ataru Tanikawa}
\email{tanikawa@ea.c.u-tokyo.ac.jp}

\author{Ataru Tanikawa}
\affiliation{Department of Earth Science and Astronomy, College of
  Arts and Sciences, The University of Tokyo, 3-8-1 Komaba, Meguro-ku,
  Tokyo 153-8902, Japan; tanikawa@ea.c.u-tokyo.ac.jp}
\affiliation{RIKEN Center for Computational Science,
  7-1-26 Minatojima-minami-machi, Chuo-ku, Kobe, Hyogo 650-0047,
  Japan}

\author{Ken'ichi Nomoto}
\affiliation{Kavli Institute for the Physics and Mathematics of the
  Universe (WPI), The University of Tokyo, 5--1--5, Kashiwanoha,
  Kashiwa, 277--8583, Japan}
%%
%%\affiliation{Hamamatsu Professor}

\author{Naohito Nakasato}
\affiliation{Department of Computer Science and Engineering,
  University of Aizu, Tsuruga Ikki-machi Aizu-Wakamatsu, Fukushima,
  965-8580, Japan}

\begin{abstract}

We study the hydrodynamics and nucleosynthesis in the
double-detonation model of Type Ia supernovae (SNe~Ia) and the
interaction between the ejecta and a surviving white dwarf (WD)
companion in the double-degenerate scenario. We set up a binary star
system with $1.0M_\odot$ and $0.6M_\odot$ carbon-oxygen (CO) WDs,
where the primary WD consists of a CO core and helium (He) shell with
$0.95$ and $0.05M_\odot$, respectively. We follow the evolution of the
binary star system from the initiation of a He detonation, ignition
and propagation of a CO detonation, and the interaction of SN ejecta
with the companion WD. The companion (or surviving) WD gets a
flung-away velocity of $\sim 1700$~km~s$^{-1}$, and captures $^{56}$Ni
of $\sim 0.03M_\odot$, and He of $3 \times 10^{-4}M_\odot$. Such He
can be detected on the surface of surviving WDs. The SN ejecta
contains a ``companion-origin stream'', and unburned materials
stripped from the companion WD ($\sim 3 \cdot 10^{-3}M_\odot$),
although the stream compositions would depend on the He shell mass of
the companion WD. The ejecta has also a velocity shift of $\sim
1000$~km~s$^{-1}$ due to the binary motion of the exploding primary
WD. These features would be prominent in nebular-phase spectra of
oxygen emission lines from the unburned materials like SN~2010lp and
iPTF14atg, and of blue- or red-shifted Fe-group emission lines from
the velocity shift like a part of sub-luminous SNe~Ia. We expect SN~Ia
counterparts to the D$^6$ model would leave these fingerprints for
SN~Ia observations.

\end{abstract}


\section{Introduction}
\label{sec:introduction}

The progenitor system of Type Ia supernovae (SNe~Ia) is one of the
biggest mysteries in astronomy and astrophysics. It is generally
thought that an SN~Ia is powered by thermonuclear explosion of a
carbon-oxygen (CO) white dwarf (WD). However, the progenitor system is
yet to be confirmed. Since a single CO~WD never starts exploding
spontaneously, an exploding CO~WD must have a companion star. The
stellar type of the companion star has been under debate. There is a
famous and long-standing dichotomy between single degenerate
\citep[SD; e.g.][]{2018SSRv..214...67N} and double degenerate
\citep[DD][]{1984ApJS...54..335I,1984ApJ...277..355W} scenarios, where
the companion star is a main-sequence or red-giant star in the SD
scenario, or is an another WD in the DD scenario. Other scenarios are
also suggested, such as the core degenerate scenario
\citep{2011MNRAS.417.1466K}.

Recent observations have revealed some significant constraints on the
SD scenario. Red-giant stars are absent in the pre-explosion images of
SN~2011fe and SN~2014J
\citep[][respectively]{2011Natur.480..348L,2014ApJ...790....3K}, which
are the closest SNe~Ia in these decades. No main-sequence star has
been detected in a supernova remnant LMC SNR 0509-67.5
\citep{2012Natur.481..164S,2017ApJ...837..111L}, although
spin-up/spin-down models can explain the non-detection
\citep{2011ApJ...730L..34J,2011ApJ...738L...1D,2012ApJ...756L...4H,2015ApJ...809L...6B}. However,
we should note some SNe~Ia indicate signals supporting the SD
scenario. For example, PTF11kx has given a signature of the
interaction of supernova (SN) ejecta and circumstellar matter
\citep{2012Sci...337..942D}, and iPTF14atg and SN~2012cg exhibit the
interaction of SN ejecta and non-degenerate companion stars
\citep[][respectively]{2015Natur.521..328C,2016ApJ...820...92M},
although for iPTF14atg and SN~2012cg these detections have been
contested
\citep[e.g.][respectively]{2016MNRAS.459.4428K,2018ApJ...855....6S}. SNe~Ia
may have several types of progenitor systems, although they may be
dominated by a single type of a progenitor system.

The DD scenario suffers from the following problem, if one assumes the
DD systems are dominant progenitor systems for
SNe~Ia. Super-Chandrasekhar DD systems, whose total mass is more than
Chandrasekhar mass, merge at a fewer rate than the SN~Ia event rate
\citep[e.g.][]{2014ARA&A..52..107M}. In the violent merger model
\citep{2010Natur.463...61P}, the primary CO~WD in a DD system is
ignited by hydrodynamical effects, and hence super-Chandrasekhar DD
systems is not necessarily needed. However,
\cite{2015ApJ...807..105S,2016ApJ...821...67S} have shown that the
violent merger model works well only when DD systems have the
companion mass of $\gtrsim 0.8 M_\odot$; the DD systems are
super-Chandrasekhar DD systems. Although
\cite{2015ApJ...800L...7K,2017ApJ...840...16K} have suggested spiral
instability after DD mergers drives thermonuclear explosions, Sato et
al's results have indicated the spiral instability can apply only to
super-Chandrasekhar DD systems. \cite{2016MNRAS.462.2486F} have
numerically demonstrated thermonuclear explosion of the primary WDs in
detached DD systems, and found that the successful DD systems are
super-Chandrasekhar DD systems. Another solution could be collisional
DD models
\citep{2009MNRAS.399L.156R,2009ApJ...705L.128R,2010MNRAS.406.2749L,2015MNRAS.454L..61D}.
\cite{2012arXiv1211.4584K} have argued DD collisions in triple systems
can account most of SNe~Ia, but it has been controversial
\citep{2014ARA&A..52..107M}.

If we take into account sub-Chandrasekhar DD systems, whose total mass
is less than the Chandrasekhar mass, the total merger rate of super-
and sub-Chandrasekhar DD systems would be comparable to the SN~Ia
event rate. Such DD systems may explode as SNe~Ia with the aid of
helium (He) ignition -- the double detonation model.

Originally, the double detonation model has been suggested as a
derivative of the SD scenario, since the companion star is a
non-degenerate star, such as a He star
\citep{1982ApJ...257..780N,1986ApJ...301..601W,1990ApJ...354L..53L,1990ApJ...361..244L}. \cite{2010ApJ...709L..64G}
and \cite{2013ApJ...770L...8P} have shown that the primary CO~WD in a
DD system can possibly experience CO detonation driven by He
detonation. In particular, a DD system in \cite{2010ApJ...709L..64G}
is a sub-Chandrasekhar DD system. The double detonation model in DD
systems requires only a small amount of He, $\lesssim 0.01M_\odot$,
since the He detonation is triggered by hydrodynamical effects of
shock compression. This model is called ``Dynamically Driven
Double-Degenerate Double-detonation (D$^6$) model'' by
\cite{2018arXiv180411163S}, or ``helium-ignited violent merger model''
by \cite{2013ApJ...770L...8P}. Hereafter, we refer to this model as
D$^6$ model for simplicity. The D$^6$ model is more advantageous than
the double detonation model in SD systems in the following
reason. Since the double detonation model in SD systems requires such
a large amount of He as $\gtrsim 0.1M_\odot$
\citep{1982ApJ...257..780N}, this model is predicted to leave behind
signature of the He detonation
\citep{1994ApJ...423..371W,2011ApJ...734...38W}; actually the
signature has been found
\citep{2017Natur.550...80J,2018ApJ...861...78M}, although the
observations of such signature has been rare.

The distinct point of the D$^6$ model from other DD models is that the
companion WD can survive thermonuclear explosion of the primary WD
\citep{2013ApJ...770L...8P}. The DD system is so close that the
surviving WD gets hypervelocity (HV) $\gtrsim 10^3$~km after the
primary WD explodes. Recently, \cite{2018arXiv180411163S} have found
out three HV~WDs from Gaia DR2. If the D$^6$ model can explain all the
SNe~Ia in the Milky Way Galaxy, one should find $\sim 30$ HV~WDs
within $1$~kpc of the Sun. The number of HV~WDs are fewer than
expected. However, if more HV~WDs would be found in near future, it
would support the statement that the D$^6$ model is a major origin of
SNe~Ia.

If the D$^6$ model would be the case for a significant fraction of
SNe~Ia, it is important to study SN ejecta and surviving WD of the
D$^6$ model. \cite{2010ApJ...709L..64G} and \cite{2013ApJ...770L...8P}
have not followed WD explosion although they have investigated the
merging process of DD systems, and He
detonation. \cite{2015MNRAS.449..942P} have focused only on
ejecta-companion interaction, manually setting up the blast wave of
SN~Ia explosion. There are several studies for the interaction of SN
ejecta with non-degenerate companions
\citep[e.g.][]{2008A&A...489..943P,2012A&A...548A...2L,2013ApJ...774...37L}.

Therefore, we numerically follow the following sequence of events: the
He detonation, CO detonation, WD explosion, and ejecta-companion
interaction by means of Smoothed Particle Hydrodynamics (SPH)
simulation coupled with nuclear reactions. Although we treat
super-Chandrasekhar DD system, we believe super- and sub-Chandrasekhar
DD systems have common features in the D$^6$ explosion.

Our paper is structured as follows. In section~\ref{sec:method}, we
present our SPH simulation method and initial conditions. In
section~\ref{sec:result}, we show simulation results. In
section~\ref{sec:discussion}, we compare our results with observations
of SNe~Ia and HV~WDs. In section~\ref{sec:summary}, we summarize this
paper.

\section{Method}
\label{sec:method}

Our SPH code is the same as used in \cite{2017ApJ...839...81T}
\citep[see
  also][]{2018MNRAS.477.3449K,2018ApJ...858...26T,2018MNRAS.475L..67T}.
We thus briefly describe our code. For equation of state (EoS), we use
Helmholtz EoS without Coulomb corrections
\citep{2000ApJS..126..501T}. We couple our SPH code with nuclear
reaction networks Aprox13 \citep{2000ApJS..129..377T}. We optimize our
code on massively parallel computing environments utilizing FDPS
\citep{2016PASJ...68...54I,2018PASJ...70...70N}, and accelerate
calculations of particle-particle interactions with AVX/AVX2/AVX512
instructions \citep[e.g.][]{2012NewA...17...82T,2013NewA...19...74T}.

Our initial condition is a binary star system consisting of
$1.0M_\odot$ and $0.6M_\odot$ WDs. The primary one has a CO core and a
He shell with $0.95M_\odot$ and $0.05M_\odot$, respectively, and the
companion one has only a CO core. In our setup, the companion WD has
no He shell. This is because we assume that its He shell transfers to
the primary WD through the prior merging process. Nevertheless, this
is an extreme case. Moreover, He shell masses on primary and companion
WDs depend on details of the He star evolution and mass transfer,
i.e. binary parameters of progenitors
\citep{1985ApJS...58..661I,1987ApJ...317..717I,1988ApJ...328..207K,1991ApJ...370..615I,2018arXiv180304444Z}.
In future, we will investigate cases where primary and companion WDs
have various total and He shell masses.

We set up the initial condition as follows. We make a single CO~WD in
the same way as \cite{2015ApJ...807...40T} \citep[see
  also][]{2015ApJ...807..105S,2016ApJ...821...67S}. We map SPH
particles consistently with the 1D profile of a fully degenerate CO~WD
with $10^6$~K, where the mass fractions of carbon and oxygen are
$X_{\rm C}=0.5$ and $X_{\rm O}=0.5$, respectively.  Subsequently, we
relax a configuration of SPH particles by evolving these SPH particles
by our SPH code. Since this relaxation involves dissipative process,
temperature of SPH particles is increased up to several $10^6$~K.  For
the primary WD, we change the CO composition of SPH particles in the
outermost shell with $0.05M_\odot$ to the mass fractions of helium,
carbon, and oxygen of $X_{\rm He}=0.6$, $X_{\rm C}=0.2$, and $X_{\rm
  O}=0.2$, respectively.  Thus, the He shell contains $0.03M_\odot$ of
helium. Note that He and C+O can be mixed in the merging process of
two WDs due to Kelvin-Helmholtz instability
\citep{2013ApJ...770L...8P}. The mixing of He and C+O facilitates
ignition and propagation of He detonation
\citep{2014ApJ...797...46S}. Since we do not smooth chemical elements
in our SPH code, we cannot follow the mixing of chemical compositions
along with the Kelvin-Helmholtz instability. This is because we do not
begin our simulation from the merging process.

We put these two WDs so that they orbit around each other on a
circular orbit with a semi-major axis of $1.6 \cdot 10^4$~km, where
the Roche-lobe radius of the companion WD is the same as its radius
according to an approximate formula of \cite{1983ApJ...268..368E}. We
assign star ID 1 and 2 to SPH particles initially belonging to the
primary and companion WDs, respectively. We put a hotspot with a size
of $10^3$~km in the He shell of the primary WD. The hotspot is located
at the orbital plane of the binary star system in the propagating
direction of the primary WD. We set such a large hotspot in order to
initiate a He detonation easily.

The total number of SPH particles is $67,108,864$. The primary and
companion WDs consist of $41,943,040$ and $25,165,824$ SPH particles,
respectively. All the particles have equal mass. This means the mass
resolution is $\sim 2.4 \cdot 10^{-8}M_\odot$. We call this mass
resolution ``fiducial mass resolution''.

For resolution check, we perform an additional simulation in the
following initial condition. We prepare the same WD as the primary WD
described above, except that the mass resolution is $2$ times
higher. Therefore, the number of SPH particles for the WD is
$83,886,080$, and the mass resolution is $\sim 1.2 \cdot
10^{-8}M_\odot$. We call this mass resolution ``higher mass
resolution''.

In section~\ref{sec:result}, we treat two coordinate systems:
Cartesian and spherical coordinate systems. In the Cartesian
coordinate system, the barycenter of the binary star system is at the
coordinate origin, and the barycenteric velocity is zero. The orbital
plane of the binary star system is set to the $x$--$y$ plane. At the
initial time, the centers of the primary and companion WDs sit on the
$x$- and $y$-axes, respectively. The angular momentum vector of the
binary star system points in the same direction of the $z$-axis. To
coordinate transformation between the Cartesian and spherical
coordinate systems, $x=r \cos\phi \sin\theta$, $y=r \sin\phi
\sin\theta$, and $z=r \cos\theta$, where $r=(x^2+y^2+z^2)^{1/2}$, and
$\theta$ and $\phi$ are the polar and azimuthal angles, respectively.

\section{Results}
\label{sec:result}

We first overview our simulation results in
section~\ref{sec:overview}. We investigate the SN ejecta and surviving
WD (or companion WD) in detail in sections~\ref{sec:snejecta} and
\ref{sec:swd}, respectively.

\subsection{Overview}
\label{sec:overview}

\begin{figure*}[ht!]
  \begin{center}
    \plotone{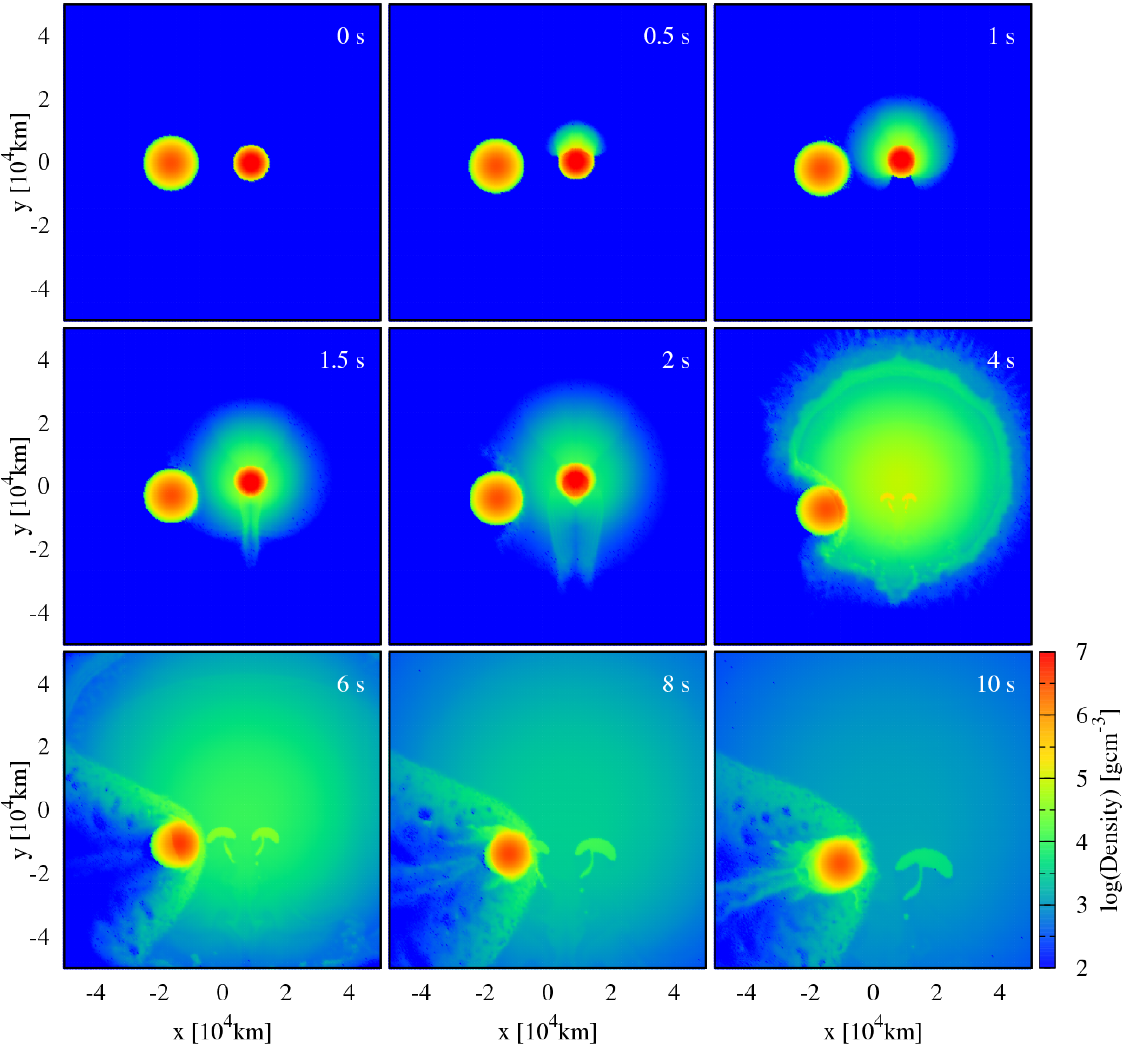}
  \end{center}
  \caption{Density distribution at $t=0$, $0.5$, $1$, $1.5$, $2$, $4$,
    $8$, and $10$~s.}
  \label{fig:viewDensityOut}
\end{figure*}

Figure~\ref{fig:viewDensityOut} shows the time evolution of the
density distribution in the binary star system. He detonation starts
in the He shell of the primary WD at the time $t=0$~s, and propagates
in the He shell, not into the CO core of the primary WD. The He
detonation converges on the back side of the primary WD from its
beginning point at $t \sim 1.25$~s. A shock wave separates from the He
detonation, invades into the CO core, and converges at an off-centered
point in the CO core just before $t=1.625$~s. Subsequently, a CO
detonation occurs at the converging point of the shock
wave. Eventually, the primary WD experiences thermonuclear explosion,
and the SN ejecta interacts with the companion WD, or to-be surviving
WD.

\begin{figure*}[ht!]
  \begin{center}
    \plotone{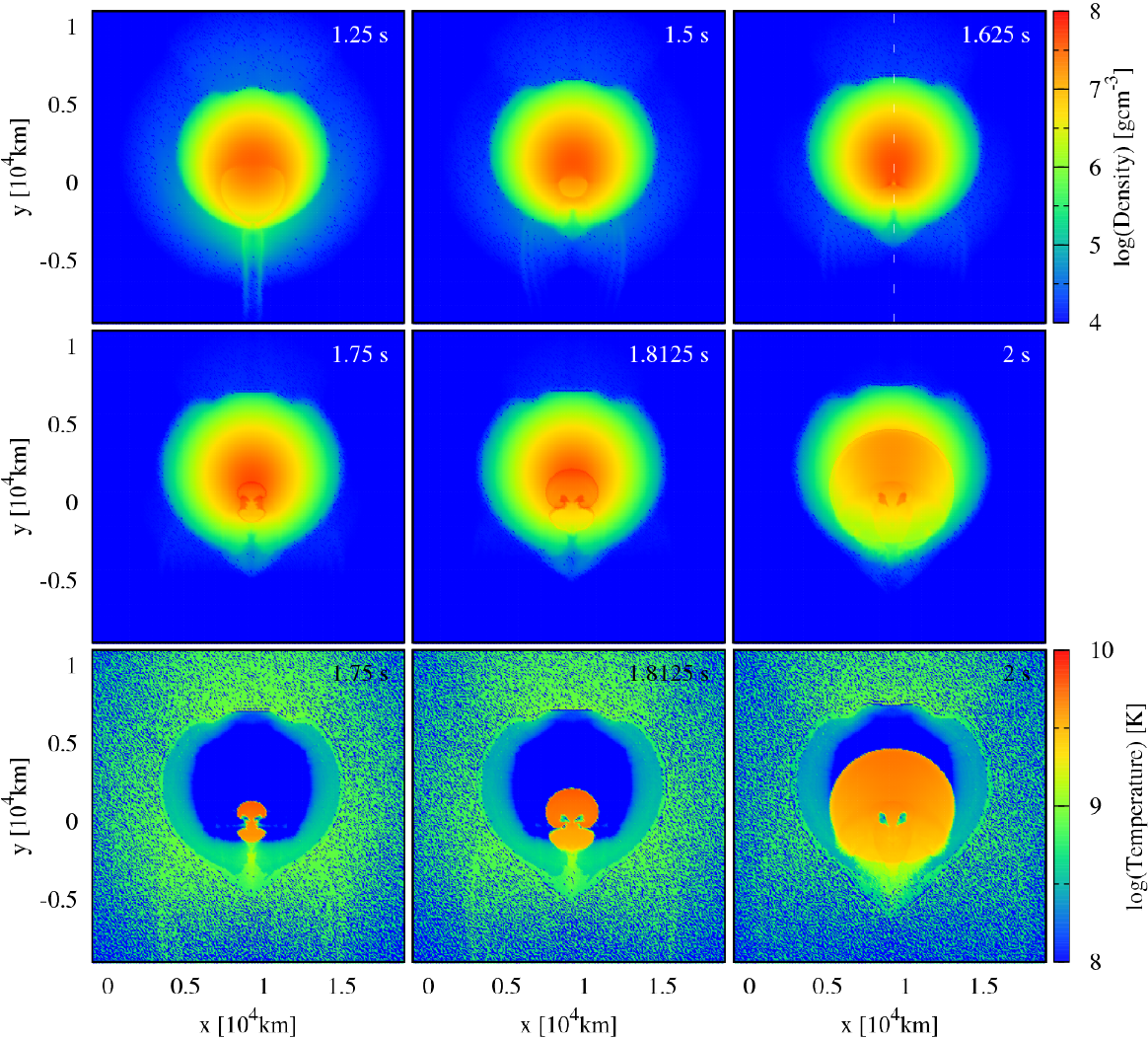}
  \end{center}
  \caption{Density distribution at $t=1.25$, $1.5$, $1.625$,
      $1.75$, $1.8125$, and $2$~s, and temperature distribution at
      $t=1.75$, $1.8125$, and $2$~s in the primary WD. The white
      dashed line in the panel at $t=1.625$~s indicates a column in
      which a shock wave invading into the CO core converges. The column
      is focused in Figure~\ref{fig:convergingShock}.}
  \label{fig:viewDensityIn}
\end{figure*}

Figure~\ref{fig:viewDensityIn} focuses on the time before and after
the shock wave converges in the CO core of the primary WD. We can see
the shock wave, the density discontinuity in the CO core, is
converging from $t=1.25$~s to $t=1.625$~s, and the CO detonation,
density (and temperature) discontinuity, is propagating from
$t=1.75$~s to $t=2$~s. The converging shock wave directly ignites the
CO detonation.

\begin{figure}[ht!]
  \plotone{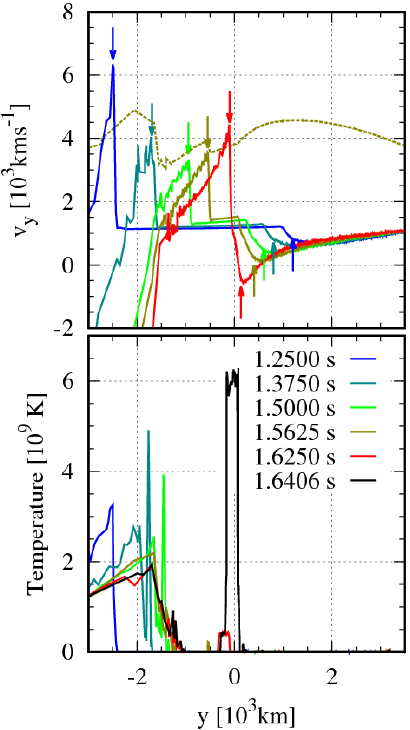}
  \caption{Profiles of $y$-velocity and temperature in a
      column in which a shock wave invading into the CO core
      converges. This column is indicated in the panel at $t=1.625$~s
      of Figure~\ref{fig:viewDensityIn}. In each panel, we draw each
      physical quantity at $t=1.25$, $1.375$, $1.5$, and $1.625$~s. We
      also show temperature at $t=1.6406$~s. In the $y$-velocity
      panel, we point to shock wave fronts by arrows, and indicate the
      sound velocity profile at $t=1.5625$~s by a dashed curve as
      reference.}
  \label{fig:convergingShock}
\end{figure}

We investigate the converging shock wave in detail.
Figure~\ref{fig:convergingShock} shows dynamics of the converging
shock wave on a line indicated by the white dashed line in
Figure~\ref{fig:viewDensityIn}. The shock wave fronts pointed by
arrows converge at $y \sim 0$~km. The front propagating in the
negative $y$-direction is less sharp than in the positive
$y$-direction, since the former front passes through a denser region,
or a region with a higher sound velocity. The shock wave front
propagates in the positive (negative) $y$-direction at an average
velocity of $\sim 6.7 \cdot 10^3$~km~s$^{-1}$ ($\sim 4.0 \cdot
10^3$~km~s$^{-1}$) from $t=1.25$~s to $t=1.625$~s, while the CO core
itself moves at a velocity of $\sim 10^3$~km~s$^{-1}$ in the positive
$y$-direction. Therefore, the velocity of the shock wave front is
$\sim 5.7$~km~s$^{-1}$ ($\sim 5.0$~km~s$^{-1}$). The corresponding
shock wave front has a velocity of $\sim 4 \cdot 10^3$~km~s$^{-1}$
($\sim 4 \cdot 10^3$~km~s$^{-1}$) in \cite[][see
  Fig.~1]{2010A&A...514A..53F}. These velocities are roughly
consistent. Since the sound velocity at the converging point ($y \sim
0$~km ) is $\sim 4 \cdot 10^3$~km~s$^{-1}$, the Mach number of the
shock wave is $\sim 1.4$ and $\sim 1.3$ in the positive and negative
$y$-directions, respectively.

We can see from Figure~\ref{fig:convergingShock} that the converging
shock wave directly ignites the CO detonation as described above. The
shock wave converges just after $t=1.625$~s (see the red curve in the
top panel of Figure~\ref{fig:convergingShock}). Temperature at the
converging point drastically rises between the times at $t=1.625$~s
and $t=1.6406$~s (see the red and black curves in the bottom panel of
Figure~\ref{fig:convergingShock}). The temperature at $t=1.6406$~s is
high enough to ignite carbon materials.

Two high-density, mushroom-shaped, regions are seen in the panels at
$t=4$ -- $10$~s. These regions have unburned CO materials. Such
unburned pockets might be formed due to low mass resolution, although
the mass resolution is quite high, $\sim 2.4 \cdot 10^{-8}M_\odot$. In
this paper, we conservatively suppose that these unburned materials
could be numerical artifacts. The reason is as follows. We focus on
the stage at $t=1.8125$~s, when the CO detonation is running in the CO
core of the primary WD. These unburned regions can be seen as
low-temperature pockets in the bottom panels at $t=1.8125$ (or $2$~s)
in Figure~\ref{fig:viewDensityIn}. These unburned regions are
surrounded by detonated regions with temperatures as high as several
$10^9$~K.  From the density and temperature panels at $t=1.8125$~s in
Figure~\ref{fig:viewDensityIn}, we find these unburned regions have
such high densities as $\sim 10^8$~g~cm$^{-3}$. Therefore, they are
unburned not due to too low densities, but due to blocking the
invasion of the CO detonation. However, a CO detonation should never
be prevented from propagating into such a high density region.

Hereafter, we do not include these unburned materials in our
discussion. For this purpose, we change their chemical compositions to
$100$~\% of $^{56}$Ni. We define these unburned materials as SPH
particles satisfying the following three conditions. (1) They
originate from the primary WD. These unburned materials come from the
primary WD as seen in Figure~\ref{fig:viewDensityIn}. (2) They are
within a distance of $3 \times 10^5$~km from the coordinate center at
$t=50$~s. We need this condition in order to avoid erasing unburned
materials in the outer SN ejecta apart from the coordinate center by
more than $5 \times 10^5$~km at $t=50$~s (see
Figure~\ref{fig:viewSnrPosition}). Note that these outer materials are
unburned for a physical reason, i.e. due to too low densities. (3)
They have the carbon mass fraction more than the critical value of
$X_{\rm C,crit}=0.2$ at $t=50$~s, since they are unburned. Then the
total mass of the unburned materials is $1.02 \cdot 10^{-2}
M_\odot$. We do not choose $X_{\rm C,crit}=0.2$ arbitrarily. In fact,
the total mass of the unburned materials is not sensitive to $X_{\rm
  C,crit}$. If we change $X_{\rm C,crit}$ to $0.1$ and $0.4$, the
total mass of the unburned materials increases and decreases by at
most $2 \cdot 10^{-6}M_\odot$, respectively.

\begin{figure}[ht!]
  \begin{center}
    \plotone{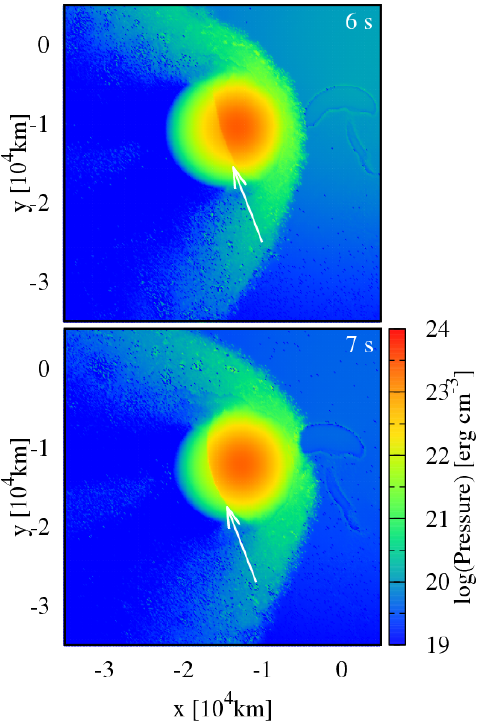}
  \end{center}
  \caption{Pressure distribution at $t=6$ and $7$~s. White arrows
    point to the pressure discontinuity.}
  \label{fig:viewPressure}
\end{figure}

\begin{figure*}[ht!]
  \begin{center}
    \plotone{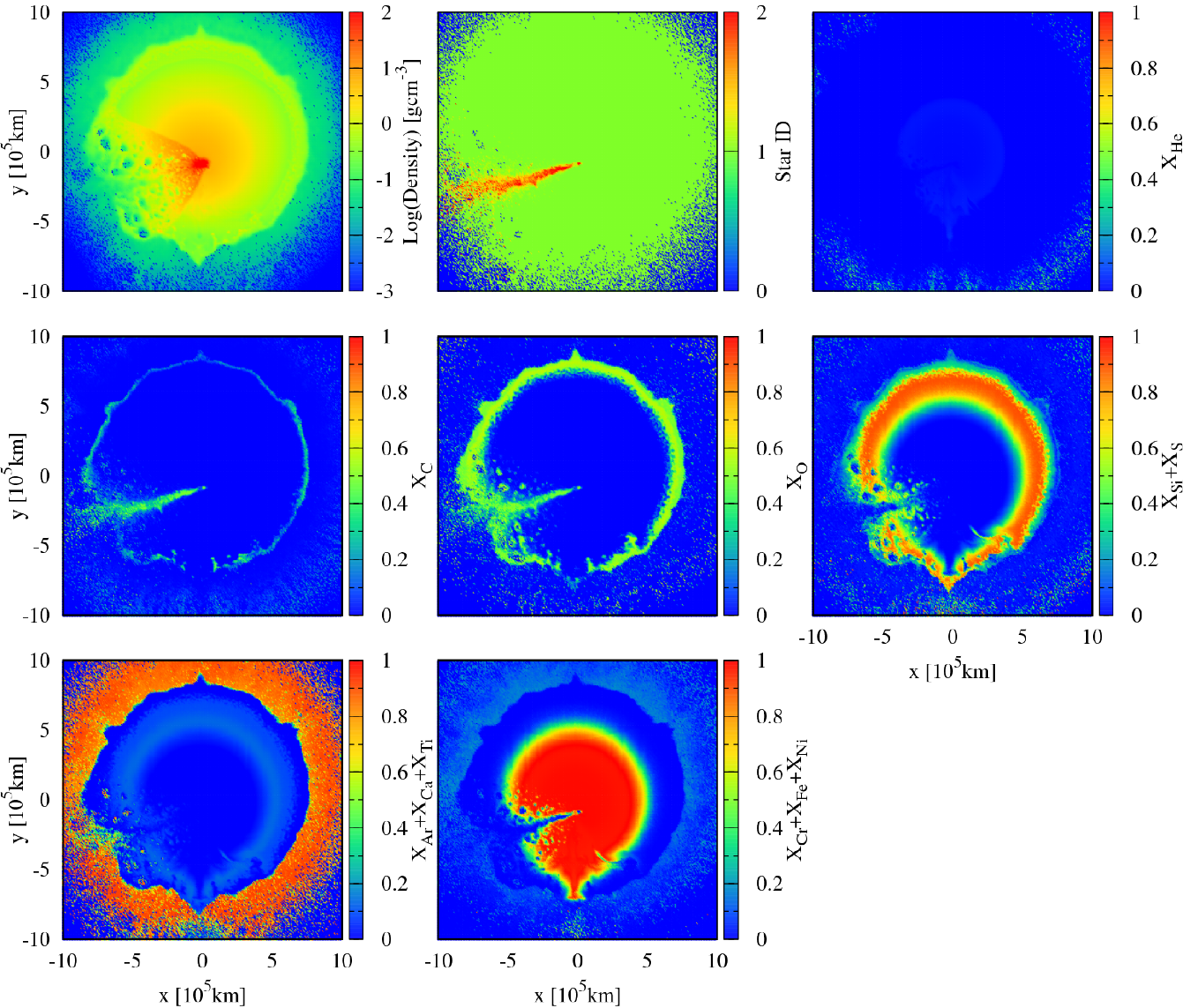}
  \end{center}
  \caption{Distribution of density, star ID, and mass fractions of
    chemical elements at $t=50$~s. Note that, if there is no material,
    the star IDs ``0'' are assigned. We change the
      mushroom-shaped, unburned materials to $100$~\% $^{56}$Ni
      materials. This is also true for
      Figures~\ref{fig:viewElement}, \ref{fig:resampleJackknife},
      \ref{fig:AngleDependent}, \ref{fig:velocityShift}, and
      \ref{fig:viewSwdProfile}.}
  \label{fig:viewSnrPosition}
\end{figure*}

The interaction between the SN ejecta and the companion WD forms an
ejecta shadow behind the companion WD (see the panels at $t=6$ --
$10$~s), which is similar to an ejecta shadow seen in
\cite{2015MNRAS.449..942P}. The interaction also strips materials of
the companion WD, which can be seen as a stream (or streams) denser
than its surroundings in the ejecta shadow at $t=8$ and
$10$~s. Hereafter, we call this stream ``companion-origin
stream''. The companion-origin stream flows out after a shock wave,
formed by collision between the SN ejecta and companion WD, passes
through the companion WD at $t \sim 8$~s. Note that the shock wave is
not formed by collision between the unburned materials and companion
WD. The shock wave can be seen as density discontinuity inside the
companion WD in the panel at $t=6$~s, and as pressure discontinuity,
pointed by white arrows, inside the companion WD in
Figure~\ref{fig:viewPressure}.

Figure~\ref{fig:viewSnrPosition} shows the distribution of density,
star ID, and mass fractions of chemical elements at $t=50$~s. Note that
they are zoomed out $20$ times compared to the panels of
Figure~\ref{fig:viewDensityOut}. In the density distribution, we can see
the ejecta shadow, circular section-shaped. The surviving (or
companion) WD is located at the vertex of the circular section. In the
star ID distribution, the surviving WD can be also found at the root
of the companion-origin stream. The stream consists of C+O.  The solid
angles of the ejecta shadow and stream are $\sim 1.8$ and $\sim 0.21$
steradians, respectively. The ejecta shadow is much wider than the
stream.

Aside from the ejecta shadow and stream, the SN ejecta has a
spherically symmetric shape. Chemical elements in the SN ejecta are
dominated by Fe-group elements (Cr, Fe, and $^{56}$Ni), the lighter
Si-group elements (Si and S), O, C, the heavier Si-group elements (Ar,
Ca, and Ti), and He in order from the inside. The heavier
silicon-group elements are the products of the He detonation in the He
shell mixed with CO compositions. Such chemical structure is typical
of the double detonation model.

As seen in Figure~\ref{fig:viewSnrPosition}, the surviving WD is
located far from the coordinate origin by several $10^4$~km, despite
that the binary system is present at the coordinate origin at
$t=0$~s. This is because the surviving WD flies away free from the
gravity of the exploding primary WD. We find the surviving WD has HV
of $\sim 1700$~km~s$^{-1}$, consistent with the velocity of the binary
motion $\sim 1800$~km~s$^{-1}$.

\subsection{Supernova ejecta}
\label{sec:snejecta}

\subsubsection{$^{56}$Ni mass}

\begin{figure}[ht!]
  \plotone{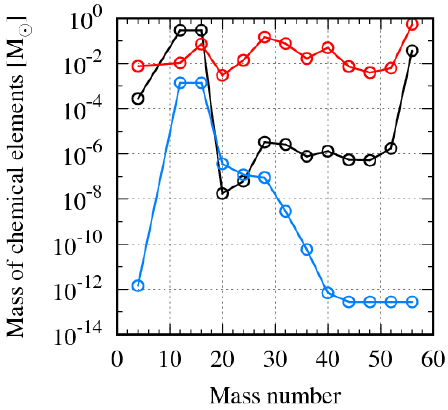}
  \caption{Masses of chemical elements in the SN ejecta (red),
    companion-origin stream (blue) and surviving WD (black) at
    $t=50$~s. The SN ejecta includes the companion-origin
    stream. Materials of the SN ejecta and surviving WD are
    gravitationally unbound and bound to the surviving WD,
    respectively.}
  \label{fig:viewElement}
\end{figure}

In Figure~\ref{fig:viewElement}, we show the mass of chemical elements
in the SN ejecta, companion-origin stream, and surviving companion
WD. The SN ejecta includes the companion-origin stream. The SN ejecta
has $\sim 0.54M_\odot$ $^{56}$Ni, most of which are synthesized by the
CO detonation in the primary WD. The He detonation yields little
$^{56}$Ni ($\sim 10^{-4} M_\odot$) because the He shell is small in
mass and contains $40$~\% (in mass fraction) of C+O which are mixed
initially.  The CO detonation also produces the lighter Si-group
elements (Si and S) of $\sim 0.22M_\odot$ through incomplete Si
burning. The unburned materials, which do not include the
mushroom-shaped materials, are $\sim 0.07M_\odot$ oxygen and $\sim
0.01M_\odot$ carbon. The He detonation synthesizes the heavier
Si-group element (Ar, Ca, and Ti) of $\sim 0.05M_\odot$, especially
dominated by Ca.

\begin{figure}[ht!]
  \plotone{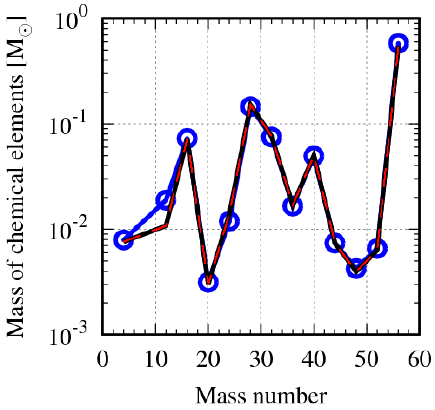}
  \caption{Mass of chemical elements for the original data
      (black), and $16$ subgroups of SN ejecta (red) from the fiducial
      mass-resolution simulation, and data from the higher
      mass-resolution simulation (blue). The black and red curves
      overlay each other, since the chemical compositions are the
      same.}
  \label{fig:resampleJackknife}
\end{figure}

Figure~\ref{fig:resampleJackknife} shows yielded chemical compositions
little depend on mass resolutions of SPH simulation as a whole. In
detail, the $^{56}$Ni mass is $\sim 0.54M_\odot$ and $\sim
0.58M_\odot$ for the fiducial and higher mass resolutions,
respectively, while the Si mass is $\sim 0.15M_\odot$ for both the
fiducial and higher mass resolutions. The mushroom-shaped, unburned
materials have mass of $\sim 0.01M_\odot$ for both the fiducial and
higher mass resolutions. The dependence of the yielded chemical
compositions on mass resolution is quite small.

In order to evaluate an error of chemical compositions of SN ejecta,
we adopt Jackknife resampling. We divide all the SPH particles into
$16$ subgroups by sampling one SPH particle per $16$ SPH particles
from SN ejecta. For each subgroups, we count masses of chemical
elements and multiply these masses by $16$. We can see from
Figure~\ref{fig:resampleJackknife} that chemical compositions are the
same among the original data and subgroups. Therefore, sampling bias
is almost zero.

We compare our nucleosynthesis yields with those in models~2 of
\cite{2010A&A...514A..53F}, which has the CO core of $0.920M_\odot$
and the He shell of $0.084M_\odot$, being similar to our primary
WD. We focus only on the products of CO detonation, since their He
shell consists of pure He. Our $^{56}$Ni mass ($\sim 0.54M_\odot$) is
larger than theirs ($0.36M_\odot$), while our Si-group mass
($0.22M_\odot$) is smaller than theirs ($0.44M_\odot$). We also
compare our products of CO detonation with those of
\cite{2011ApJ...734...38W}.  In their models with the primary WD of
$\sim 1M_\odot$, $^{56}$Ni masses are $0.5M_\odot$ --
$0.8M_\odot$. Our nucleosynthesis yields are roughly consistent with
those in previous studies. However, further detailed nucleosynthesis
studies are necessary to reach a better agreement.

\subsubsection{Nucleosynthesis yields in the velocity space}

The total mass of the companion-origin stream is $\sim 3 \cdot
10^{-3}M_\odot$. It consists of mostly $50$~\% carbon and $50$~\%
oxygen in mass, being almost the same as the original compositions,
but includes a small amount of Ne, Mg, and Si, $\sim 10^{-7}M_\odot$
for each. They are synthesized by shock heating when the SN ejecta
collides with the companion WD.

\begin{figure*}[ht!]
  \plotone{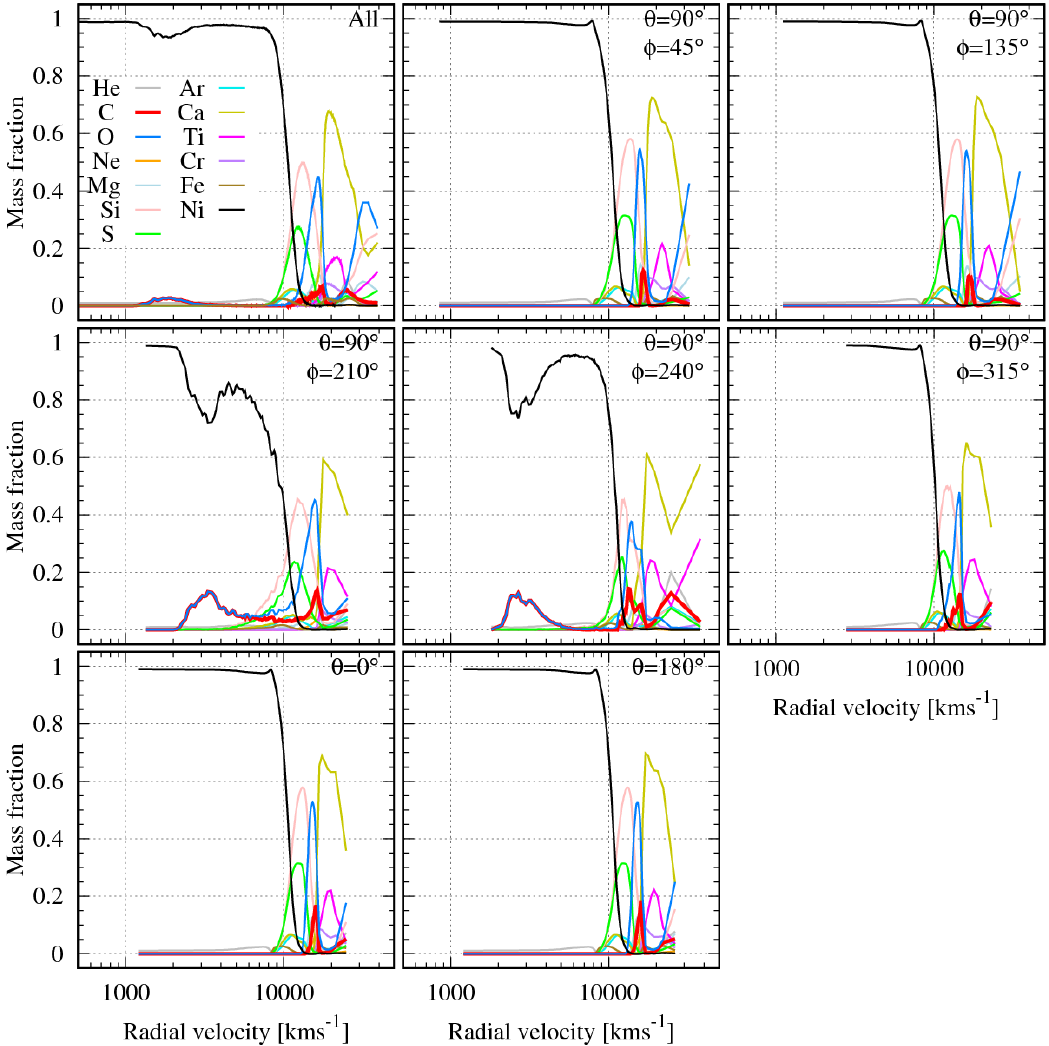}
  \caption{Chemical elements in mass fraction as a function
    of radial velocity.}
  \label{fig:AngleDependent}
\end{figure*}

Figure~\ref{fig:AngleDependent} shows the chemical elements in mass as
a function of the radial velocity. We average the mass fractions over
all the angle, and find that the abundance structure is similar to the
typical double detonation model. The first high-velocity components
($\gtrsim 2$ -- $3 \cdot 10^4$~km~s$^{-1}$) consist of Ca and Ti,
synthesized by the He detonation. The second high-velocity components
($\sim 1.5$ -- $2 \cdot 10^4$~km~s$^{-1}$) are composed of C+O,
unburned materials located at the outer region in the CO core of the
primary WD. Behind the unburned materials, the third high-velocity
components ($\sim 1$ -- $1.5 \cdot 10^4$~km~s$^{-1}$) includes Si and
S, products of incomplete silicon burning. Low-velocity components ($<
10^4$~km~s$^{-1}$) are composed of $^{56}$Ni. The low-velocity
components contain a slight amount of C+O at the velocity of $\sim 3
\cdot 10^3$~km~s$^{-1}$, described below in detail.

The low-velocity C+O come from the companion-origin stream. The
features show up from the view into which the stream comes,
$(\theta,\phi)=(90^\circ,210^\circ)$ and
$(90^\circ,240^\circ)$. Interestingly, the chemical structure from
these views is the same as the averaging structure at the velocity of
$> 10^4$~km~s$^{-1}$. Therefore, C+O have bimodality in the velocity
distribution.  One component has a higher velocity than Si and S
($\sim 1.5$ -- $ 2 \cdot 10^4$~km~s$^{-1}$), and the other has a lower
velocity ($\sim 3 \cdot 10^3$~km~s$^{-1}$). Moreover, the low-velocity
C+O have a velocity dependent on the viewing angle, $\gtrsim 3 \cdot
10^3$~km~s$^{-1}$ from the view of
$(\theta,\phi)=(90^\circ,210^\circ)$, and $\lesssim 3 \cdot
10^3$~km~s$^{-1}$ from the view of
$(\theta,\phi)=(90^\circ,240^\circ)$.

We emphasize that the low-velocity C+O originate from the
companion-origin stream, not from the unburned materials of the
primary WD. As seen in the star-ID panel of
Figure~\ref{fig:viewSnrPosition}, the companion-origin stream comes
into sight from the viewing angles
$(\theta,\phi)=(90^\circ,210^\circ)$ and
$(90^\circ,240^\circ)$. Moreover, we change the mushroomed-shaped,
unburned materials to materials $100$~\% $^{56}$Ni as described above.

The distribution of C+O in our SN ejecta is different from the delayed
detonation model \citep{2013MNRAS.429.1156S,2018ApJ...861..143L},
although both models have low-velocity C+O. In the delayed detonation
model, C+O are extended from the high-velocity to low-velocity
components. On the other hand, C+O in our SN ejecta have two peaks in
the velocity space.

No low-velocity C+O can be seen from other views $\theta=0^\circ$ and
$180^\circ$, and $(\theta,\phi)=(90^\circ,45^\circ)$,
$(90^\circ,135^\circ)$, and $(90^\circ,315^\circ)$, since the stream
does not come into these sights. The velocity distributions of
chemical elements from these viewing angles are the same as those in
the double detonation model, such that materials are dominated by the
He-detonation products (Ca and Ti), unburned materials (CO
compositions), incomplete silicon burning products (Si and S), and
$^{56}$Ni in the descending order of the velocity.

\begin{figure*}[ht!]
  \plotone{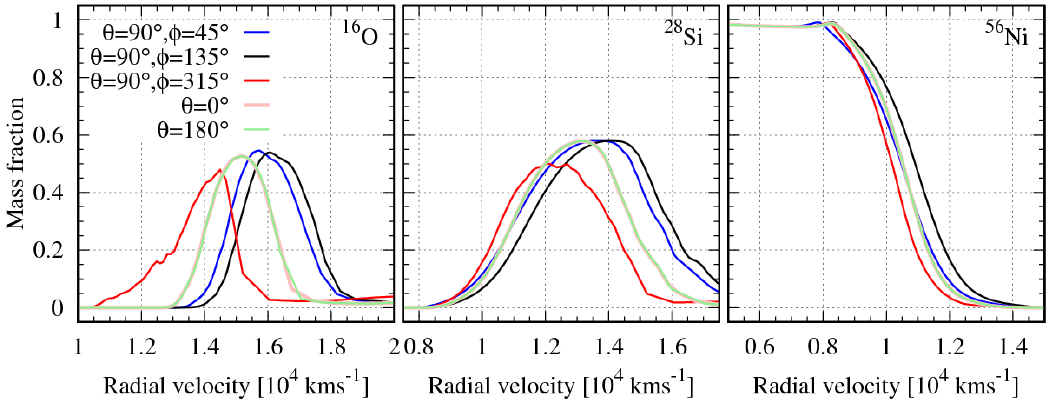}
  \caption{Mass fractions of oxygen, silicon, and nickel as a function
    of the radial velocity.}
  \label{fig:velocityShift}
\end{figure*}

The surviving WD moves at a speed of $1700$~km~s$^{-1}$ in the
bottom-right direction (closely to
$(\theta,\phi)=(90^\circ,315^\circ)$) in
Figure~\ref{fig:viewSnrPosition}, not only at $t=50$~s but also at the
explosion time of the primary WD. Hence, the primary WD also
propagates at a speed of $\sim 1100$~km~s$^{-1}$ in the opposite
direction (closely to $(\theta,\phi)=(90^\circ,135^\circ)$) at the
explosion time. Therefore, the SN ejecta should be shifted in the
propagating direction of the exploding primary
WD. 

Figure~\ref{fig:velocityShift} shows the velocity distribution of O,
Si, and $^{56}$Ni observed from various viewing angles. All chemical
elements have higher velocities for the viewing angles closer to the
direction of the bulk velocity of the SN ejecta. The velocity
difference is $\sim 1$ -- $2 \cdot 10^3$~km~s$^{-1}$ between
$(\theta,\phi)=(90^\circ,135^\circ)$ and
$(\theta,\phi)=(90^\circ,315^\circ)$. The velocity of these elements
observed from the other viewing angles is intermediate between the
velocity observed from the two viewing angles. This is consistent with
the velocity of the binary motion of the exploding primary WD,
$1100$~km~s$^{-1}$.

The velocity difference does not come from the asymmetric explosion of
the double detonation model. In the asymmetric explosion, when the
velocity of O and Si from a viewing angle is larger than from another
viewing angle, the velocity of $^{56}$Ni from the former is smaller
than from the latter \citep[see Fig.~6
  in][]{2010A&A...514A..53F}. Note that the bulk motion of the
exploding primary WD (or SN ejecta) systematically increases all the
velocity of O, Si, and $^{56}$Ni observed from the viewing angle in
the propagating direction.

\subsection{Surviving white dwarf}
\label{sec:swd}

The surviving WD has the total mass of $\sim 0.6M_\odot$, roughly
equal to the initial mass. However, it captures a small amount of
materials originally from the exploding primary WD. The total mass of
the captured materials is $\sim 0.03M_\odot$, dominated by $^{56}$Ni
of $\sim 0.03M_\odot$ (or $^{56}$Ni of $\sim 0.02M_\odot$ unless we
change the mushroomed-shaped, unburned materials to materials of
$100$~\% $^{56}$Ni). The captured materials contain slight amount of
helium ($\sim 3 \cdot 10^{-4} M_\odot$). The captured materials
consist of $^{56}$Ni and He in the following reason. They are captured
due to their low velocity. Hence, they are located at the center of
the explosion, i.e. in a high-density region. In general, when CO
detonation passes a high-density region, it mostly synthesizes
$^{56}$Ni owing to the rapid nuclear reactions, and leaves a small
amount of He as residuals of He produced by photo-dissociation.
\cite{2017ApJ...834..180S} have estimated that a $0.6M_\odot$ WD
captures $0.03M_\odot$ $^{56}$Ni, which is consistent with our
results.

\begin{figure*}[ht!]
  \plotone{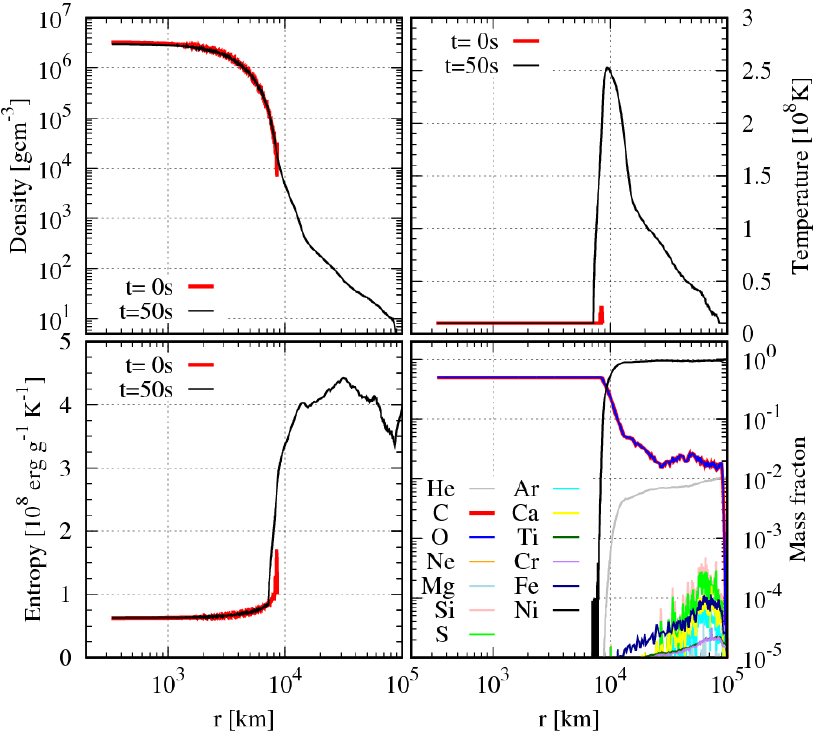}
  \caption{1D profiles of density, temperature, and entropy of the
    companion WD at $t=0$~s (red) and $t=50$~s (black), and of
    chemical elements of the companion WD at $t=50$~s.}
  \label{fig:viewSwdProfile}
\end{figure*}

We investigate 1D profiles of the surviving WD at $t=50$~s shown in
Figure~\ref{fig:viewSwdProfile}, where we count only gravitationally
bound materials to the surviving WD. We find its internal structure
(at $r \lesssim 7 \cdot 10^{3}$~km) is virtually undamaged, comparing
its density, temperature, and entropy profiles at $t=0$ and
$50$~s. Moreover, the surviving WD keeps its C+O.

On the other hand, its external structure is changed by the
interaction with the SN ejecta. The surviving WD captures a part of
the SN ejecta, and gets an envelope consisting of $^{56}$Ni, C, O, and
He in descending order in mass. These $^{56}$Ni and He result from
nucleosynthesis in the primary WD, while C and O come from the
surviving WD. Note again that we change the mushroom-shaped, unburned
materials to materials of $100$~\% $^{56}$Ni. The C and O originally
from the surviving WD are modestly stripped from the surviving WD by
the SN ejecta. The envelope has high temperature ($\sim 2.5 \cdot
10^8$~K) and entropy ($\sim 4.5 \cdot 10^8$~erg~g$^{-1}$~K$^{-1}$) due
to nuclear reactions in the primary WD, and due to the shock heating
arising from the collision between the SN ejecta and surviving WD. The
entropy of the captured materials is slightly higher than estimated by
\cite{2017ApJ...834..180S}, $1$ -- $3 \cdot
10^8$~erg~g$^{-1}$~K$^{-1}$. Hence, the captured materials are
slightly less bound than their estimate.

\section{Discussion}
\label{sec:discussion}

\subsection{Exploding primary white dwarf}

First, we discuss SN~Ia counterparts to the D$^6$ model, based on the
results shown in section~\ref{sec:result}. Here, we ignore products
yielded by the He detonation, which are the heavy Si-group elements
(Ar, Ca, and Ti) with such high velocities, $\gtrsim 2 \cdot
10^4$~km~s$^{-1}$. This reason is as follows. We set the He shell of
the primary WD to be so thick ($\sim 0.05M_\odot$) that the He and CO
detonation easily occurs in our simulation. However, the D$^6$ model
would succeed when the He shell is $\lesssim 0.01M_\odot$
\citep{2010ApJ...709L..64G,2013ApJ...770L...8P}, and would indicate
smaller signal of these products than our simulation results.

The most prominent feature in our SN ejecta is the companion-origin
stream (see Figure~\ref{fig:viewSnrPosition}). Owing to the presence
of the stream, the abundance of unburned materials has two peaks in
velocity space from specific viewing angles, as seen in
Figure~\ref{fig:AngleDependent}. The lower velocity component of the
unburned materials (a few $10^3$~km~s$^{-1}$) can be observed as
oxygen emission lines in nebular-phase spectra. Such oxygen emission
lines have been observed in SN~2002cx-likes
\citep{2006AJ....132..189J,2007PASP..119..360P}, and a part of
SN~2002es-like SN~2010lp
\citep{2013ApJ...775L..43T,2013ApJ...778L..18K} and iPTF14atg
\citep{2016MNRAS.459.4428K}. In SN~2002cx-likes, possibly explained by
pure-deflagration explosion \citep{2005A&A...437..983K} \citep[but
  see][]{2013A&A...559A..94W}, $^{56}$Ni prevails from the inner to
outer ejecta, while our SN ejecta confines $^{56}$Ni to the inner
parts with $\lesssim 10^4$~km~s$^{-1}$. Hence, we rule out
SN~2002cx-likes as D$^6$ explosion candidates.

SN~2010lp and iPTF14atg could be promising counterparts to D$^6$
model, since their light curves and spectral evolutions are consistent
with the explosion of sub-Chandrasekhar mass WDs
\citep[][respectively]{2013ApJ...778L..18K,2016MNRAS.459.4428K}. Although
iPTF14atg have ultraviolet (UV) pulse due to collision of the SN
ejecta with the non-degenerate companion \citep{2015Natur.521..328C},
the UV pulse could be explained by surface radio activity of $^{56}$Ni
produced by the He detonation \citep{2016MNRAS.459.4428K}. However,
our SN ejecta may be inconsistent with SN~2010lp. SN~2010lp has both
blue- and red-shifted oxygen emissions in its nebular spectra. On the
other hand, our SN ejecta would have either of blue- or red-shifted
oxygen emissions, since the companion-origin ejecta stream propagates
in one direction from the explosion center. Note that it may be
difficult to identify these oxygen emissions, since the
companion-origin stream has small mass ($\sim 3 \cdot
10^{-3}M_\odot$). We need to study nebular-phase spectra of the D$^6$
model by performing radiative transfer calculations
\cite[e.g.][]{2010ApJ...708.1703M,2017ApJ...845..176B}.

We should note that these oxygen feature may not be observed in the
following reason. The companion WD should have a He shell in reality,
although it does not have in our setup. If the He shell has more than
$\sim 3 \cdot 10^{-3}M_\odot$, the SN ejecta may strip only the He
shell, not the CO core. Therefore, the companion-origin stream can
consist of He materials. Nevertheless, the companion-origin stream can
contain CO because of mixing of CO into the overlying He shell during
common envelope phase and merging process as mentioned for the He
shell of a primary WD. Even though the He shell has more than $\sim 3
\cdot 10^{-3}M_\odot$, C+O is likely mixed in the companion-origin
stream. Moreover, a companion WD with larger mass has smaller He shell
mass. Eventually, the presence and absence of oxygen futures depends
on detail binary parameters of progenitor systems. In future, we will
investigate compositions of companion-origin streams in the cases of
companion WDs with different total and He shell masses.

Another feature is the velocity shift of SN ejecta due to the binary
motion of the primary WD, $\sim
10^3$~km~s$^{-1}$. \cite{2011MNRAS.413.3075M} have shown iron and
nickel emission lines can be tracers of such a velocity shift.
\cite{2018MNRAS.tmpL.101D} have compiled cobalt emissions in nebular
spectra of various SNe~Ia, and have found the cobalt emissions are
both blue- and red-shifted in SNe~Ia with $-19<M_{\rm B}<-18$
(SN~2007on, SN~2003hv, and SN~2003gs), and either blue- or red-shifted
in those with $M_{\rm B}>-18$ (SN~2016brx, SN~2005ke, SN~1999by, and
SN~1991bg). Although they have attributed these blue- and red-shifted
features to the collisional DD model
\citep{1989ApJ...342..986B,2010MNRAS.406.2749L,2009MNRAS.399L.156R,2010ApJ...724..111R,2009ApJ...705L.128R,2012ApJ...759...39H,2015MNRAS.454L..61D},
SNe~Ia with either of blue- or red-shifted Fe-group emissions can be
also explained by the D$^6$ model.

\subsection{Surviving white dwarf companion}

Hereafter, we describe issues related to the surviving WD.
\cite{2017ApJ...834..180S} have discussed post-supernova winds blown
by radioactive $^{56}$Ni on the surfaces of surviving WDs. We can
compare our results with their surviving CO~WD model with
$0.6M_\odot$. They have modeled the surface of the surviving CO~WD,
such that the mass of radioactive $^{56}$Ni is $0.0003$ --
$0.03M_\odot$, and the entropy of its surface is $1$ -- $3 \cdot
10^8$~erg~g$^{-1}$~K$^{-1}$. As we obtain the $^{56}$Ni mass and
entropy on the surface of the surviving WD to be $\sim 0.03M_\odot$,
and $4.5 \cdot 10^8$~erg~g$^{-1}$~K$^{-1}$, our simulation results are
consistent with their modeling, although materials on the surface in
our results are slightly less bound than those in their models. Thus,
SN~2011fe could not be explained by the D$^6$ model, which is the same
conclusion as theirs. This is because SN~2011fe would be more luminous
than observed if it contained a surviving WD.

We discuss the surface abundance of the surviving WD. First, we
consider the surface pollution by interstellar medium (ISM) and
interstellar objects (ISOs). The surviving WD could accrete ISM
through the Bondi-Hoyle-Lyttleton accretion. The accreting mass is
estimated as
\begin{align}
  M_{\rm acc} &\sim \dot{M}_{\rm acc,BHL} T_{\rm disk} \nonumber \\ 
  &\sim \frac{\pi \rho_{\rm ism} G^2 M_{\rm wd}^2}{c_{\rm s,ism}^4}
  \frac{h_{\rm disk}}{v_{\rm wd}} \nonumber \\
  &\sim 1.0 \cdot 10^{20}
  \left( \frac{n_{\rm ism}}{1 \mbox{cm}^{3}} \right)
  \left( \frac{c_{\rm s,ism}}{20\mbox{km~s}^{-1}} \right)^{-4}
  \left( \frac{h_{\rm disk}}{200\mbox{pc}} \right)
  \nonumber \\
  &\times \left( \frac{M_{\rm wd}}{0.6M_\odot} \right)
  \left( \frac{v_{\rm wd}}{2000 \mbox{km~s}^{-1}} \right)^{-1} \;
  \mbox{[g]},
\end{align}
where $\dot{M}_{\rm acc,BHL}$ is the mass accretion rate through
Bondi-Hoyle-Lyttleton accretion, $T_{\rm disk}$ is time the surviving
WD spending in the Galactic disk, $\rho_{\rm ism}$, $n_{\rm ism}$, and
$c_{\rm s,ims}$ are, respectively, ISM mass density, number density,
and sound speed, $h_{\rm disk}$ is the scale height of the Galactic
disk, and $M_{\rm wd}$ and $v_{\rm wd}$ are the mass and velocity of
the surviving WD. Note that this estimate constrains on the upper
limit of the accreting mass \citep{2006ApJ...638..369K}.  Moreover, we
estimate a collision rate of the surviving WD with ISOs like
1l/`Oumuamua \citep{2017Natur.552..378M}. The estimate method is the
same as in \cite{2018PASJ...70...80T}. Then, the surviving WD collides
with ISOs at most once, and accrete the ISO mass of $\sim 10^{13}$~g
at most. Eventually, the surviving WD accretes ISM and ISO mass much
less than materials captured from the SN ejecta by several orders of
magnitude. Hence, ISM and ISOs cannot pollute the surface of the
surviving WD.

As shown in section~\ref{sec:swd}, the surviving WD captures $^{56}$Ni
of $\sim 0.03M_\odot$, and He of $\sim 3 \cdot 10^{-4}M_\odot$. The
$^{56}$Ni of $\sim 0.03M_\odot$ actually includes the mushroom-shaped,
unburned materials of $\sim 0.01M_\odot$. Even if the unburned
materials are not numerical artifacts, they cannot be regarded as
anomaly, since the surviving WD also has similar unburned materials.
The $^{56}$Ni will undergo radioactive decay. The $^{56}$Ni decay
products could be identified as anomalous abundances. However, the
decay products do not necessarily stay on the surface of the surviving
WD, since they will receive sedimentation
\citep{1986ApJS...61..197P,1992ApJS...82..505D}. Note that they can
keep their position due to radiative levitation
\citep{1995ApJS...99..189C,1995ApJ...454..429C}. It must be necessary
to perform sophisticated numerical calculation to follow the time
evolution of the surviving WD if we know whether the decay products
stay on the surface of the surviving WD. Here, we do not perform such
calculations.

The surviving WD certainly has He on its surface, since He does not
experience sedimentation. However, it is difficult to assess whether a
HV~WD is a surviving WD against the D$^6$ explosion on the basis of
the detection of He in the following two reasons. First, the detection
of He on a HV~WD cannot be the smoking-gun evidence that the HV~WD is
a surviving WD against the D$^6$ explosion. Since WDs generally have
He on its surface, a HV~WD gets its HV through mechanism other than
the D$^6$ explosion. Second, the non-detection of He on a HV~WD's
surface does not always deny the HV~WD is a D$^6$ candidate. He on a
HV~WD's surface can be seen only when the HV~WD has high temperature
on its surface. In fact, \cite{2018arXiv180411163S} did not found He
on the surface of their HV~WDs (WD1) for this reason. In summary, we
can say a HV~WD is not a surviving WD against the D$^6$ explosion only
if He is not detected despite of high temperature on the HV~WD's
surface. We reemphasize WD1 in \cite{2018arXiv180411163S} can be a
surviving WD against the D$^6$ explosion despite of the non-detection
of He, since WD1's surface has low temperature.

Since LP~40--365 (or GD~492) has high abundance of Mn and other iron
group elements, it is thought to be a WD candidate surviving against
the Type Iax explosion
\citep{2017Sci...357..680V,2018ApJ...858....3R,2018MNRAS.479L..96R}.
Such abundance pattern could not be reconciled with the D$^6$
explosion, since the D$^6$ explosion involves sub-Chandrasekhar mass
WD.

\section{Summary}
\label{sec:summary}

In order to study features of SN ejecta and surviving WD in the D$^6$
model, we perform SPH simulation of a binary star system with
$1.0M_\odot$ and $0.6M_\odot$ CO~WDs, where the primary WD has a He
shell with $0.05M_\odot$ mixed with C+O. The primary WD undergoes
thermonuclear explosion following the He detonation on the shell and
the CO detonation in the core. The SN ejecta collides with the
companion WD, and the interaction of the SN ejecta with the companion
WD form the ejecta shadow and companion-origin
stream. \cite{2015MNRAS.449..942P} have also found out such ejecta
shadows in their simulations for their D$^6$ models. The companion WD
survives the explosion of the primary WD, and flies away at velocity
of $\sim 1700$~km~s$^{-1}$ as the surviving WD.

The SN ejecta has typical features of the double detonation explosion
on average. However, there are two different features from the double
detonation explosion. (1) First, the SN ejecta strips materials of the
companion WD, and contains the companion-origin ejecta consisting of
C+O. The companion-origin ejecta can make oxygen emission lines in
nebular-phase spectra. Therefore, SN~Ia counterparts to the D$^6$
model can be a part of SN~2002es-likes, such as SN~2010lp and
iPTF14atg which have oxygen emission lines in their nebular-phase
spectra. Note that the compositions of the companion-origin ejecta may
depend on the He shell mass of the companion WD. In future, we will
investigate this dependence. (2) Second, the SN ejecta has velocity
shift of $\sim 1000$~km~s$^{-1}$ due to the binary motion of the
exploding primary WD. This velocity shift can result in blue- or
red-shifted Fe-group emission lines in nebular-phase spectra seen in
sub-luminous SNe~Ia, such as SN~2016brx, SN~2005ke, SN~1999by, and
SN~1991bg.

The surviving WD certainly has He on its surface. The He originates
from residuals of He produced by photo-dissociation at the center of
the primary WD. However, since WDs generally have He on their
surfaces, the presence of He could not be the smoking-gun evidence of
surviving WDs against the D$^6$ explosion. The surviving WD also has
$^{56}$Ni decay products on its surface just after it survives the
explosion of the primary WD. However, the decay products would
experience sedimentation and radiative levitation. In order to
determine the surface abundance of the surviving WD, we should follow
the long-term evolution of the surviving WD.

Finally, we summarize observational features of SNe~Ia under the D$^6$
explosion. At an early time, its light curve may show a UV pulse due
to radioactive nuclei yielded by the He detonation. At the
maximum-light time, its spectra indicate Si absorption lines similarly
to ordinary SNe~Ia. At late times, in the nebular-phase, oxygen
emission lines can be observed, where the oxygen originates from the
companion-origin stream stripped by the SN ejecta. From specific
viewing angles, blue- or red-shifted Fe-group emission lines can be
also seen due to the binary motion of the exploding primary WD.

\acknowledgments

Numerical computations were carried out on Oakforest-PACS at Joint
Center for Advanced High Performance Computing, and on Cray XC50 at
Center for Computational Astrophysics, National Astronomical
Observatory of Japan. The software used in this work was in part
developed by the DOE NNSA-ASC OASCR Flash Center at the University of
Chicago. This research has been supported by World Premier
International Research Center Initiative (WPI Initiative), MEXT,
Japan, by the Endowed Research Unit (Dark side of the Universe) by
Hamamatsu Photonics K.K., by MEXT program for the Development and
Improvement for the Next Generation Ultra High-Speed Computer System
under its Subsidies for Operating the Specific Advanced Large Research
Facilities, by ``Joint Usage/Research Center for Interdisciplinary
Large-scale Information Infrastructures'' and ``High Performance
Computing Infrastructure'' in Japan (Project ID: jh180021-NAJ), and by
Grants-in-Aid for Scientific Research (16K17656, 17K05382, 17H06360)
from the Japan Society for the Promotion of Science.

\software{Modules of Helmholtz EoS and Aprox13 in FLASH
  \citep{2000ApJS..131..273F,2010ascl.soft10082F}}

%\bibliography{natbib}

\end{document}